\documentclass[reprint]{revtex4-1}

\usepackage{scrextend}

\usepackage{amsmath}
\usepackage{amssymb}
\usepackage{graphicx}
\usepackage{dcolumn}
\usepackage{bm}
\usepackage{color}
\usepackage[usenames,dvipsnames]{xcolor}

\usepackage{bm}
\usepackage{graphicx}
\usepackage{ulem}

\def\be{\begin{equation}}
\def\ee{\end{equation}}
\def\ber{\begin{eqnarray}}
\def\eer{\end{eqnarray}}
\def\bwt{\begin{widetext}}
\def\ewt{\end{widetext}}

\begin{document}

\title {Coulomb drag between a carbon nanotube and monolayer graphene} 
\author{S. M. Badalyan}
\affiliation{Center for Graphene Research, Frisco, Texas 75035, USA} 
\author{A. P. Jauho}
\affiliation{Center for Nanostructured Graphene,  Technical University of Denmark, 2800 Kongens Lyngby, Denmark}

\date{\today}

\begin{abstract}
We study Coulomb drag in a system consisting of a carbon nanotube (CNT) and monolayer graphene. Within the Fermi liquid theory we calculate the drag resistivity and find that the dimensional mismatch of the system components leads to a dependence of the drag rate on the carrier density, temperature, and spacing, which is substantially different from what is known for graphene double layers.
Due to the competing effects of forward and backward scattering, we identify new features of the drag dependence on the electron density, which allows us to control their relative contribution to the drag resistivity.
\end{abstract}

\pacs{71.45.Gm; 73.20.Mf; 73.21.Ac; 81.05.ue}


\maketitle

\paragraph{Introduction}

Coulomb drag in double well systems has been of considerable theoretical and experimental interest for several decades. The emergence of graphene has significantly expanded the physical regimes in drag experiments~\cite{Levchenko2015}. In particular, the insulating barrier between the subsystems can have a thickness down to a few atomic layers, which makes the interaction phenomena more pronounced and gives access to physics~\cite{Dean2016,Tutuc2016,PhKim2017,PhKim201717,Dean2017} which is unattainable in semiconductor samples.

Many recent works address Coulomb drag in dimensionally symmetric graphene based structures: drag between two graphene layers has been studied both experimentally and theoretically~\cite{Tutuc2011,Ponomarenko2012,Ponomarenko2013,Tse2007,Katsnelson2011,Hwang2011,Narozhny2012,Carrega2012,SMB2012}. Coupled one-dimensional systems have been also considered~\cite{Lunde2005,Shylau2014}.
%
%
Clearly the drag properties depend essentially on the system dimensionality. One may thus expect that a dimensional {\it mismatch} of the electronic subsystems can significantly affect Coulomb correlations and characteristics of the drag resistance. Until recently, however, drag between dimensionally mismatched subsystems has attracted less attention and has been limited to a few theoretical works on conventional 1D--2D systems~\cite{Sirenko1992,Lyo2003}. 

The recent experimental realization of graphene-based dimensionally mismatched electronic structures between a carbon nanotube (CNT) and a graphene monolayer~\cite{Kim2016,Kim201819} acts as an excellent stimulus for further experimental and theoretical work in this interesting direction.
The plasmon spectrum in systems of Coulomb coupled graphene nanoribbon and monolayer graphene has been recently calculated~\cite{SMB2017,Hwang2018}.
It is predicted that due to the strong interlayer Coulomb coupling, these hybrid systems behave effectively as one dimensional and do not support 2D plasmon modes with a square-root dispersion~\cite{SMB2017}. 

In the present paper, we study Coulomb drag in dimensionally mismatched graphene systems, consisting of either a metallic or a semiconducting CNT and monolayer graphene. Adopting Fermi liquid theory, we calculate the dependence of the drag resistivity on the carrier density, temperature, and spacing between the CNT and graphene.
We find that the screening effect, taken into account within the random phase approximation (RPA), strongly suppresses the drag rate and qualitatively changes its dependence on the system parameters.
The dimensional mismatch leads to a dependence of the drag resistivity on the carrier density, temperature, and spacing, which differs substantially from that known for symmetric 2D-2D or 1D-1D electronic systems. Meantime, the temperature and spacing dependence is found to be rather close to the behavior obtained in Ref.~\onlinecite{Lyo2003} for conventional 1D-2D systems.
We also show that the transresistivity for systems with a semiconducting CNT exhibits a slight dip or upturn depending, respectively, on the carrier density in a CNT or graphene, at densities corresponding to the matched Fermi wave vectors. This is because the 2D momentum is not conserved in this hybrid system and the backscattering events, which are, in general, possible in semiconducting CNTs, are suppressed due to the presence of graphene. Thus, these distinctive features in the density dependence of the drag resistivity allow us to distinguish and tune the backward and forward scattering contributions to 1D-2D drag.

\begin{figure}[t]
\includegraphics[width=.7\columnwidth]{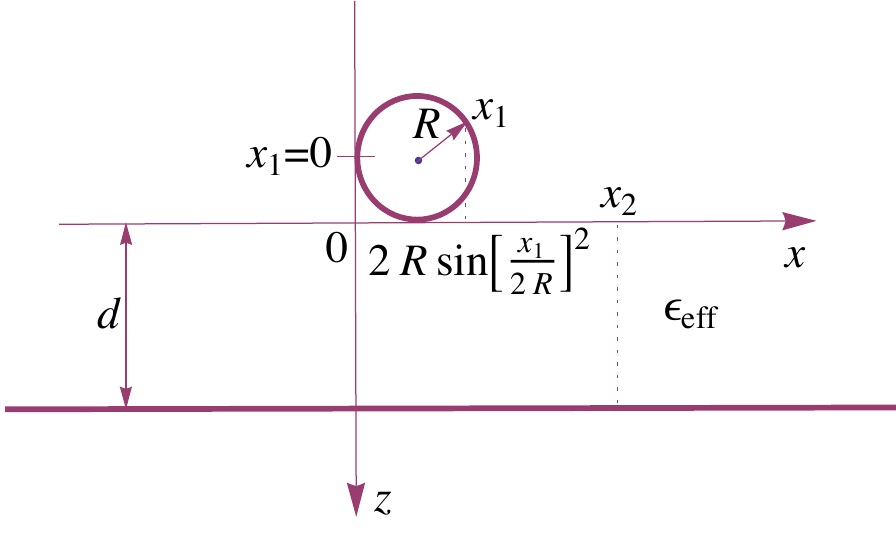}
\caption{Schematic view of a dimensionally mismatched graphene system. The bold lines show the $x-z$ profile of Coulomb coupled carbon nanotube (metallic or semiconducting) of the radius $R$ and monolayer graphene, separated by a barrier with the $\epsilon_\text{eff}$ dielectric constant and of the thickness $d$ along the $z$ direction. The $0\leq x_1\leq 2\pi R$ and $-\infty < x_2< +\infty$ coordinates correspond, respectively, to the electron positions on the cylindrical surface of CNT and on the plane of graphene.}
\label{fig1}
\end{figure}

\paragraph{Theoretical model}

We consider the following geometry ({\it cf.} Fig.~\ref{fig1}). A carbon nanotube of radius $R$ is separated by a barrier of thickness $d$ from a graphene monolayer.  The system has translational invariance only along the $y$ direction.
The electronic states in the CNT are described by the quantum numbers $n,\nu,s,k$, where the transverse quantization subband index is an integer, $n=0,\pm 1,\dots$; $\nu$ refers to the type of CNT ($\nu$=0 refers to metallic nanotubes, both armchair and zigzag, while $\nu=\pm 1$ correspond to semiconducting zigzag nanotubes).
The chirality index $s$ describes the conduction ($s = +1$) and valence ($s = -1$) bands, and $k$ is the conserved wave vector along the $y$-direction. The single-particle energy spectrum in CNT $\varepsilon^{n,\nu}_\text{1D}(s,k)=s \hbar v_{\text{gr}} \sqrt{k^2_{n,\nu}+k^2}$ where $k_{n,\nu}=\left(n-\frac{\nu}{3}\right)/R$ and $v_{\text{gr}}$ is the electron velocity in graphene~\cite{Ando2005}.
In monolayer graphene the quantum numbers $(s,{\vec p})$ describe the 2D electron spinor states in the $(x,y)$ plane with the in-plane momentum ${\vec p}$ and the single-particle Dirac spectrum $\varepsilon_\text{2D}\left(s, |{\vec p}|\right) = s v_{\text{gr}} p$.
In the present work we assume spin and valley degeneracy with the degeneracy factors $g_s=2$ and $g_v=2$ (interlayer Coulomb interaction is small for intervalley electronic transitions because of the large values of the transferred momentum) and restrict ourselves to the consideration of low temperatures and low levels of doping.

In a metallic (semiconducting) CNT with a 1D carrier density $n_\text{1D}\equiv\frac{N}{L}=2\pi R n_\text{2D} \approx 1.9\times 10^6$ cm$^{-1}$, corresponding to the areal density $n_\text{2D}\equiv\frac{N}{L^2}=3\times10^{12}$ cm$^{-2}$ in graphene and to the tube radius $R=1$ nm, we find the 1D Fermi energy $\varepsilon^\text{1D}_{\text{F}} \approx 1128.8$ K ($250.7$ K). Here $N$ is the number of carriers in CNT or graphene.
Instead, for $n_\text{2D}=3\times10^{11}$ cm$^{-2}$ taking the Fermi energies in metallic (semiconducting) CNT and graphene equal, $\varepsilon^\text{1D}_{\text{F}}=\varepsilon^\text{2D}_{\text{F}} = 740.2$ K, we have $n_\text{1D}\approx 1.24\times 10^6$ cm$^{-1}$ ($3.24 \times 10^6$ cm$^{-1}$). Note that $\varepsilon^\text{1D}_{\text{F}}\propto n_\text{1D}$ or $ n^2_\text{1D}$ and $\varepsilon^\text{2D}_{\text{F}}\propto \sqrt{n_\text{2D}}$, respectively, in a metallic or semiconducting CNT and in monolayer graphene.
On the other hand, the lowest inter-subband gap between the $n=+1$ and $n=0$ energy levels is $\Delta\varepsilon^0_\text{1D}=\hbar v_{\text{gr}}/R\approx 7624.5$ K in metallic CNT while $\Delta\varepsilon^1_\text{1D}=\hbar v_{\text{gr}}/3R\approx 2541.5$ K in semiconducting CNT. Thus, with these values the Fermi energies and temperatures are smaller than the transverse quantization energy, $\varepsilon^\text{1D}_{\text{F}}, T\ll \Delta\varepsilon^\nu_\text{1D}$, so that electronic transitions to the higher energy subbands do not make a significant contribution to drag. We adopt this lowest subband approximation in CNT and thus omit the subband index $n$.

\paragraph{Drag resistance}

The drag resistance $R_D$ in the CNT-graphene systems can be measured in two different configurations. In the first one, a current $I_G$ flowing through the graphene monolayer induces a voltage $V_\text{CNT}$ along the nanotube. Then, assuming that the normalization length, $L$, is the same along the wire and in both $x$ and $y$ directions in graphene, the drag resistance $R^\text{2D--1D}_\text{D}=V_\text{CNT}/I_\text{G}=\rho_\text{2D--1D}$ is given just by the transresistivity $\rho_\text{2D--1D}$. In the second configuration, the roles of CNT and graphene are switched, and the transresistance is determined as $R^\text{1D--2D}_\text{D}=V_\text{G}/I_\text{CNT}=\rho_\text{1D--2D} L/2\pi R$. Thus, because of the asymmetry of the CNT-graphene system the drag {\it resistivity} is different in the two configurations, however the drag {\it resistance} measured along the wire or across graphene obeys the Onsager reciprocal relation and is independent of the choice of the configuration: $R_\text{D}\equiv R^\text{1D--2D}_\text{D}= R^\text{2D--1D}_\text{D}$.

We shall assume that the electronic system can be described as a Fermi liquid both in graphene and in the CNT.
The electrical current in the system is restricted by impurity scattering and we adopt the Boltzmann equation approach~\cite{JS1992}, treating interlayer interaction perturbatively.
From the balance of the carrier distribution due to the external electric field and scattering events, we find the drag resistivity as
\begin{eqnarray}\label{DR}
\rho^\nu_\text{2D--1D}&=&\frac{h}{e^2}\frac{1}{n_\text{1D}n_\text{2D}  T}\frac{1}{L^2} \sum_{\vec q} \int d\hbar\omega
\frac{v_{12}\left(q\right)^2 \left| I\left(q_x\right)\right|^2}{\left|\epsilon_{\text{1D-2D}}\left({\vec q}, \omega\right)\right|^2}
\nonumber \\
&\times &
\frac{\Gamma^{\nu}_\text{1D}\left(q_y,\omega\right) \Gamma_\text{2D}\left({\vec q},\omega\right)}{\sinh\left(\hbar \omega/2T\right)^2}~.
\end{eqnarray}
Here $v_{12}\left(q\right)=2\pi e^2 e^{-q d} / q \epsilon_\text{eff} $ is the 2D Fourier transform of the bare interlayer Coulomb interaction with $q=\sqrt{q_x^2+q_y^2}$ and $\epsilon_\text{eff}$ is the effective low frequency dielectric function of the insulating barrier. The form factor $I\left(q_x\right)=e^{i q_x R} J_0(q_x R)$ with $J_0(x)$  the Bessel function of the first kind. We assume that $d\gg R$.

The dynamical screening function of the hybrid 1D-2D electronic system within the random phase approximation is given~\cite{SMB2017,Lyo2003} by $\epsilon_{\text{1D-2D}}\left({\vec q},\omega \right)=\epsilon^\text{eff}_{\text{1D}}\left(q_y,\omega \right)\epsilon_{\text{2D}}\left(q,\omega \right)$ where
\begin{eqnarray}\label{EPS}
\epsilon^\text{eff}_{\text{1D}}\left(q_y,\omega \right)&=& \epsilon_\text{1D}\left(q_y, \omega \right)
-
{\cal Q}_{\text{1D-2D}}\left(q_y,\omega \right) \Pi_{1 \text{D}}\left(q_y, \omega \right)
\end{eqnarray}
with
\begin{eqnarray}\label{Q}
{\cal Q}_{\text{1D-2D}}\left(q_y,\omega \right)=\frac{1}{L} \sum _{q_x} \frac{\left|I\left(q_x\right)\right|^2
v_{12}\left(q\right)^{2}\Pi_{2 \text{D}}\left(q, \omega \right)}{\epsilon_{2 \text{D}}\left(q, \omega \right)} ~.
\end{eqnarray}
The intralayer dynamical screening functions (the Lindhard polarization functions) in 1D~\cite{Que2002,Lunde2005,Shylau2014,SMB2015} and 2D~\cite{Guinea2006,Sarma2007,Kotov2012} electronic systems are, respectively, denoted by $\varepsilon_\text{1D}\left(q_y, \omega \right)$ ($\Pi_{1 \text{D}}\left(q_y, \omega \right)$) and $\varepsilon_\text{2D}\left(q, \omega \right)$ ($\Pi_{2\text{D}}\left(q, \omega \right)$).


The nonlinear response function in the CNT~\cite{Lunde2005} is
\begin{eqnarray}\label{Gamma1D}
&&\Gamma^\nu_\text{1D}\left(q_y,\omega\right)=\frac{e}{\pi\hbar \mu_\text{1D}}\frac{1}{L} \sum _{s,s';k,k'}  \delta_{k',k+q_y} F^{\nu}_\text{1D}\left(s,k;s',k'\right) \nonumber\\
&\times&
\Im \frac{h^\nu_\text{1D}\left(s,k;s',k'\right)
\left[f\left(\varepsilon^\nu_\text{1D}(s',k')\right)-f\left(\varepsilon^\nu_\text{1D}(s,k)\right)\right]}
{\varepsilon^\nu_\text{1D}\left(s,k\right)-\varepsilon^\nu_\text{1D}\left(s',k'\right)+\hbar\omega+i0^+} 
\end{eqnarray}
while for graphene~\cite{Narozhny2012,Carrega2012} it is given by
\begin{eqnarray}\label{Gamma2D}
&&\Gamma^y_\text{2D}\left(\vec q,\omega\right)=\frac{e}{\pi\hbar \mu_\text{2D}}\frac{1}{L^2} \sum_{s,s';\vec p, \vec p'}\delta_{\vec p',\vec p+\vec q}
F_\text{2D}\left(s,\vec p;s',\vec p'\right) \nonumber\\
&\times&
\Im \frac{h^y_\text{2D}\left(s, p; s', p'\right)
\left[f\left(\varepsilon_\text{2D}(s', p')\right)-f\left(\varepsilon_\text{2D}(s, p)\right)\right]}
{\varepsilon_\text{2D}\left(s, p \right)-\varepsilon_\text{2D}\left(s', p' \right)+\hbar\omega+i0^+}~.
\end{eqnarray}
Here the spinor overlap factors, stemming from the Coulomb matrix elements, are given in CNT (graphene) by $F^{\nu}_\text{1D}\left(s,k;s',k'\right)=\left(1+s s' \cos\theta^\nu_{k k'}\right)/2$ ($F_\text{2D}\left(s,\vec p;s',\vec p'\right)=\left(1+s s' \cos \theta_{\vec p \vec p'}\right)/2$) with $\theta^\nu_{k k'}$ ($\theta_{\vec p \vec p'}$) the angle between the vectors $(k_\nu, k)$ and $\left(k_\nu, k'\right)$ ($\vec p$ and $\vec p'$).
The Fermi functions are given by $f(\varepsilon)$.
We introduce also the functions $h^\nu_\text{1D}\left(s,k;s',k'\right)=\tau^\nu_\text{1D}\left(k'\right)v^\nu_\text{1D}\left(s',k'\right)-\tau^\nu_\text{1D}\left(k\right)v^\nu_\text{1D}\left(s,k\right)$
and
$h^y_\text{2D}\left(s,p;s',p'\right)=\tau_\text{2D}\left(p'\right)v^y_\text{2D}\left(s', p'\right)-\tau_\text{2D}\left(p\right)v^y_\text{2D}\left(s, p\right)$ for CNT and graphene.  Respectively, the carrier mobilities and velocities are $\mu_\text{1D}$, $v^\nu_\text{1D}\left(s,k\right)=\partial_k \varepsilon^\nu_\text{1D}\left(s,k\right)$ and $\mu_\text{2D}$, $v^y_\text{2D}\left(s, p\right)=\partial_{p_y} \varepsilon_\text{2D}\left(s, p\right)$.
In contrast to the mobility (a quantity averaged over the carrier energy), the momentum relaxation transport time $\tau(k)$ is a momentum dependent quantity, which is linear in the energy for the dominant type of disorder scattering of Dirac electrons in graphene~\cite{DasSarma2011}.
It has been shown, however, that the energy dependence does not affect calculations of the nonlinear susceptibility in graphene~\cite{Narozhny2012,Carrega2012} and semiconducting CNT~\cite{Shylau2014}; therefore we evaluate $\tau^{\pm1}_\text{1D}\left(k\right)=\tau^{1}_\text{1D}$ and $\tau_\text{2D}\left(p\right)=\tau_\text{2D}$ at the Fermi level and view them as constants.
However, this approximation is not justified for a metallic CNT where the function $h^0_\text{1D}\left(s,k;s',k'\right)$ with the constant relaxation time vanishes for forward scattering events. Therefore, in this case we include the energy dependence of the momentum relaxation time, and assuming the relaxation time is linear in the energy~\cite{Hwang2011},  $\tau^{0}_\text{1D}\left(k\right)={\tilde \tau}^{0}_\text{1D} |k|$, we find $h^0_\text{1D}\left(s,k;s',k'\right)={\tilde \tau}^0_\text{1D} v_\text{gr} \left(s' k'- s k\right)$ for a metallic CNT.

\begin{figure}[t]
\includegraphics[width=.49 \columnwidth]{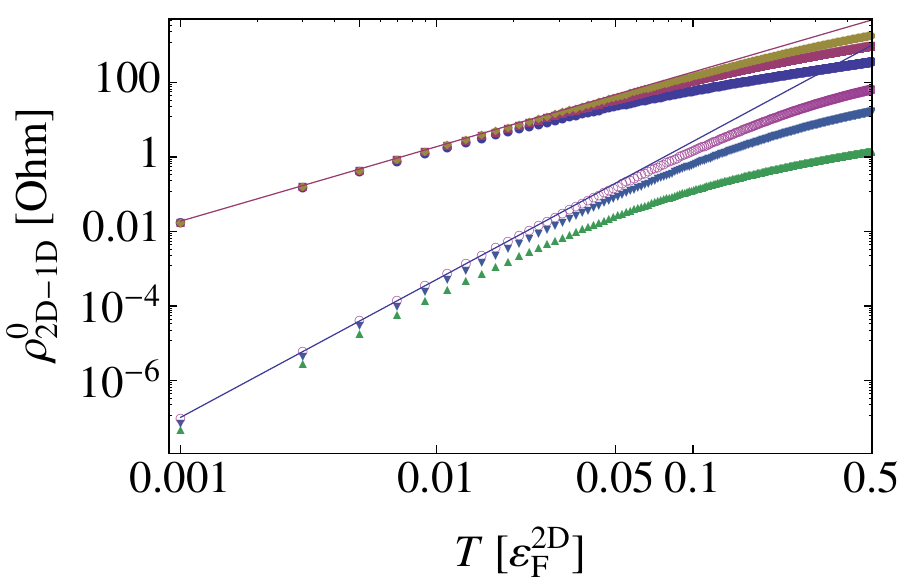}
\includegraphics[width=.49 \columnwidth]{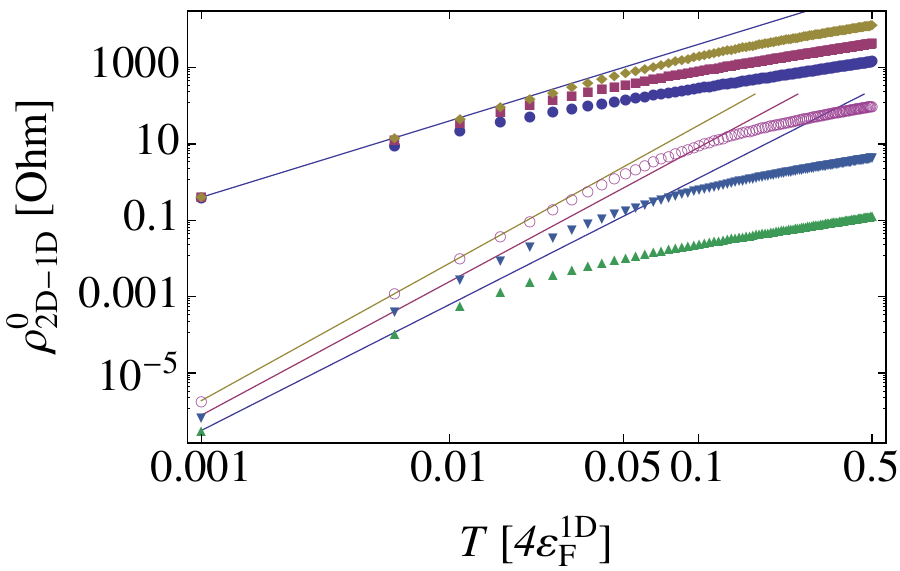}
\caption{The temperature dependence of the drag resistivity between a metallic (left) or semiconducting (right) CNT and graphene. Symbols show the transresistivity in the log-log scale with (down wisp) and without (up wisp) including the screening effect for spacing $d=30$, $10$, and $3$ nm (down up). The radius of CNT $R=1$ nm and the carrier densities $n_\text{2D}=3\times10^{11}$ cm$^{-2}$ and $n_\text{1D}=1.24\times10^{6}$ cm$^{-1}$ ($n_\text{2D}=2.7\times10^{12}$ cm$^{-2}$ and $n_\text{1D}=3.71\times10^{6}$ cm$^{-1}$) in a metallic (semiconducting) CNT--graphene hybrid system. The solid thin lines represents the $T^\beta$ power law behavior as a guide to the eye with $\beta=2$ (up wisp) and $\beta=3.7$ (down wisp) on the left panel and with $\beta=2$ (up wisp) and $\beta=3.3, 3.5$, and $3.6$ (down up, down wisp) on the right panel. }
\label{fig3}
\end{figure}

\paragraph{Low temperature regime}

From this on we restrict our discussion to the low temperature regime, $T\ll \varepsilon^\text{1D}_{\text{F}}, \varepsilon^\text{2D}_{\text{F}}$. In this case only electronic transitions within the $s=s'=1$ band contribute to the  nonlinear response functions and they can be calculated analytically. 
In particuar, for a metallic CNT, we find
\begin{eqnarray}\label{Gamma1DMLT}
\Gamma^0_\text{1D}\left(q_y,\omega\right)=\frac{e {\tilde \tau}^{0}_\text{1D}}{\hbar \mu_\text{1D}}
\frac{\omega^2}{2\pi \hbar v_\text{gr}}
\left[\delta\left(\omega+v_\text{gr}q_y\right)-\delta\left(\omega-v_\text{gr}q_y\right)\right]~.
\end{eqnarray}
%
%
%
The screening function in Eqs.~(\ref{DR})-(\ref{Q}) can be approximated in the static limit, $\epsilon_\text{1D--2D}\left(\vec q,0\right)$. 
In graphene we use the static polarizability  $\Pi_\text{2D}\left(q, 0 \right)= -2 k^\text{2D}_\text{F}/\pi \hbar v_\text{F}$ for $q<2k^\text{2D}_\text{F}$~\cite{Guinea2006} and we have $\epsilon_\text{2D}\left(q, 0 \right)= 1+4\alpha_\text{gr} k^\text{2D}_\text{F}/q$ with $\alpha_\text{gr} = e^2/\hbar v_\text{F} \epsilon_\text{eff}$. 
In numerical calculations we take $\epsilon_\text{eff}=4$ mimicking a hBN barrier.
In a metallic CNT, the static polarizability is approximated as $\Pi_\text{1D}\left(q_y, 0 \right)= -1/\pi^2 \hbar v_\text{F}$~\cite{Que2002,Lunde2005} while the bare interaction in $\epsilon_\text{1D}\left(q, 0 \right)$ is $v_\text{1D}\left(q \right)=2\hbar v_\text{F} \alpha_\text{gr} I_0(|q_y| R)K_0(|q_y| R)$~\cite{Ando2005}, and we find $\epsilon^\text{eff}_\text{1D}\left(q_y, 0\right)$ as a one-dimensional integral and calculate it numerically. Here $I_0(y)$ and $K_0(y)$ are the modified Bessel functions of the first and second kind.
In a semiconducting CNT the static polarizability $\Pi_\text{1D}\left(q_y, 0 \right)$, according to its definition, is represented as an additional one-dimensional integral in $\epsilon^\text{eff}_\text{1D}\left(q_y, 0\right)$.

\begin{figure}[t]
\includegraphics[width=.49\columnwidth]{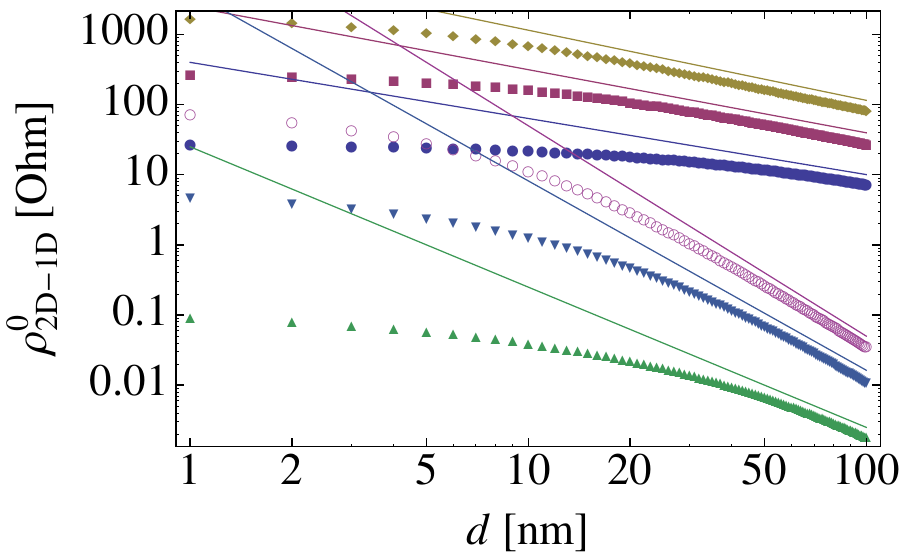}
\includegraphics[width=.49\columnwidth]{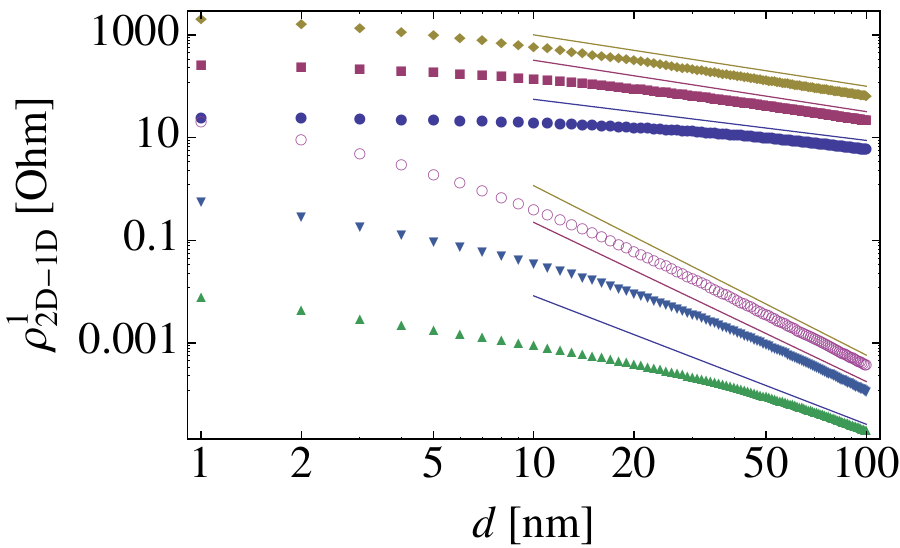}
\caption{The drag resistivity between a metallic (left) or semiconducting (right) CNT and graphene as a function of spacing $d$. Symbols show the transresistivity in the log-log scale with (down wisp) and without (up wisp) including the screening effect at different temperatures $T=30$, $100$, and $300$ K (down up).
The other parameters are the same as in Fig.~\ref{fig3}.
%
The solid thin lines represents the $d^{-\beta}$ behavior as a guide to the eye with $\beta=2, 2.7, 3; 0.8, 0.9, 1$ and $\beta=2.5, 3.1, 3.3; 0.8, 1, 1$ (down up), respectively, on the left and right panels.
}
\label{fig4}
\end{figure}

\paragraph{Numerical results and discussion}

We first discuss the temperature dependence of the drag resistivity between a CNT and graphene. As seen in Fig.~\ref{fig3}, the transresistivity without screening shows approximately the familiar $T^2$-dependence, which originates from the interplay between the 1D--2D phase space behavior in drag scattering events at low $T$, and the long wavelength singularity of the unscreened interaction. In contrast to drag in conventional 2D systems~\cite{JS1992}, here the integrations over $q_x$, $q_y$, and $\omega$ are not decoupled into products of one-dimensional integrals. Therefore, the static screening effect, along with the strong suppression of the absolute drag resistivity, changes qualitatively the drag behavior as a function of $T$ and substantially enhances this dependence ({\it cf.} Fig.~\ref{fig3}).
We observe a similar effect also for the interlayer spacing dependence of drag. In the absence of screening the overall weak dependence on $d$ ({\it cf.} Fig.~\ref{fig4}) is due to the long wavelength singularity of the unscreened interaction, which is much stronger in this hybrid 1D--2D system than in 1D--1D electronic systems. Even after screening is turned on, the drag resistivity remains a weakly decreasing function with $d$ for small values of $d\lesssim 10$ nm. For relatively large values of $d\gtrsim 50$ nm, the decrease of the drag resistivity becomes rather strong and can be fitted by a power law function $d^{-\beta}$ with $\beta\sim 3$ at $T\gtrsim 300$. The index $\beta$ decreases with a decrease of $T$.

\begin{figure}[t]
\includegraphics[width=.49\columnwidth]{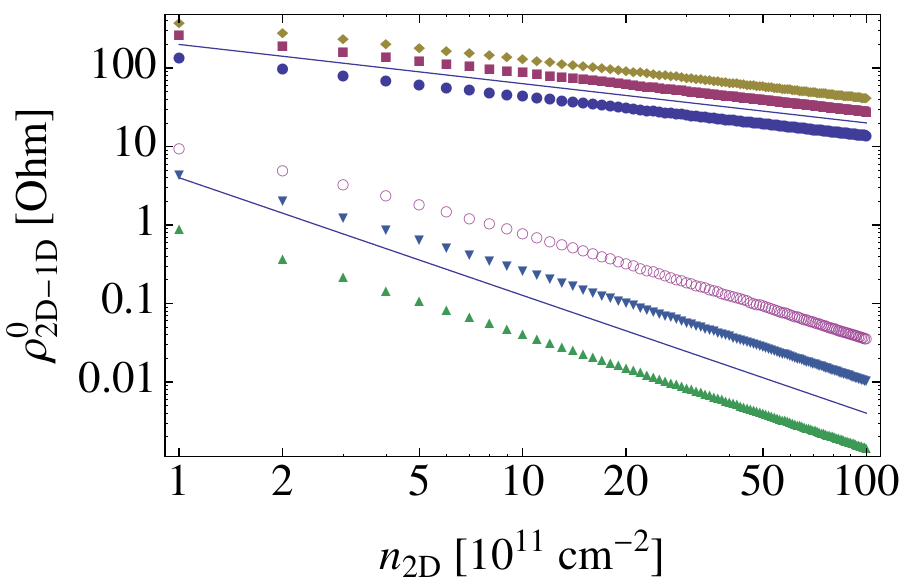}
\includegraphics[width=.49\columnwidth]{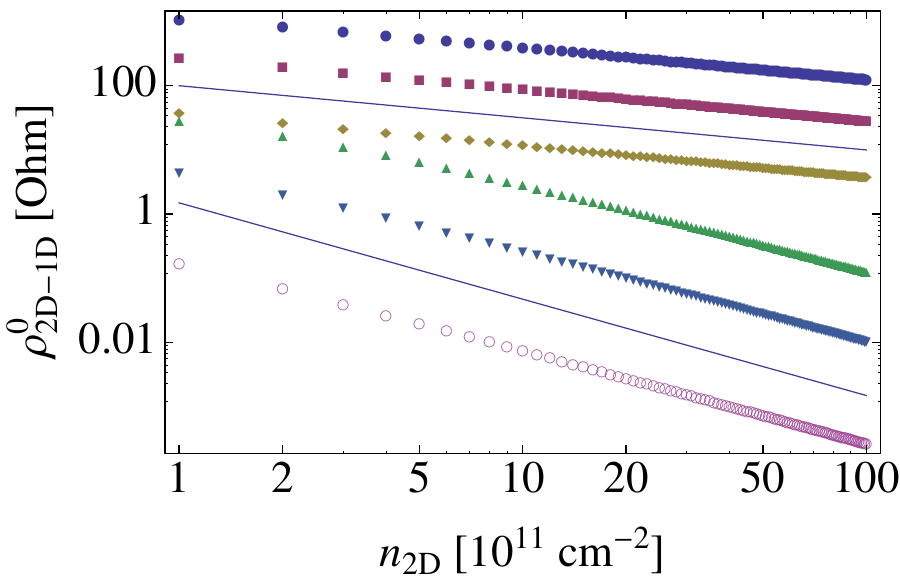}
\caption{The dependence of the drag resistivity between a metallic CNT and graphene on the carrier density in graphene for $d=30$, $10$, and $3$ nm (down up) at $T=100$ K (left) and at $T=30$, $100$, and $300$ K (down up) for $d=10$ nm (right). In each panel the down (up) wisp of symbols show log-log plots of the transresistivity with (without) including the screening effect. The radius of CNT is $R=1$ nm and the carrier density $n_\text{1D}=1.24\times10^{6}$ cm$^{-1}$. The solid thin lines represent the $n_\text{2D}^{-0.5}$ (up) and $n_\text{2D}^{-1.5}$ (down) behaviors as a guide to the eye.}
\label{fig5}
\end{figure}

\begin{figure}[t]
\includegraphics[width=.49\columnwidth]{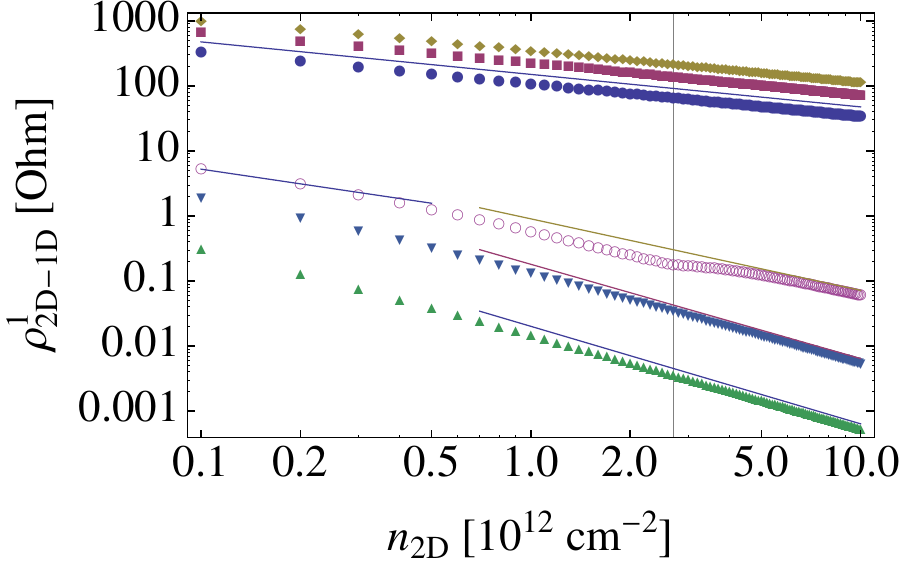}
\includegraphics[width=.49\columnwidth]{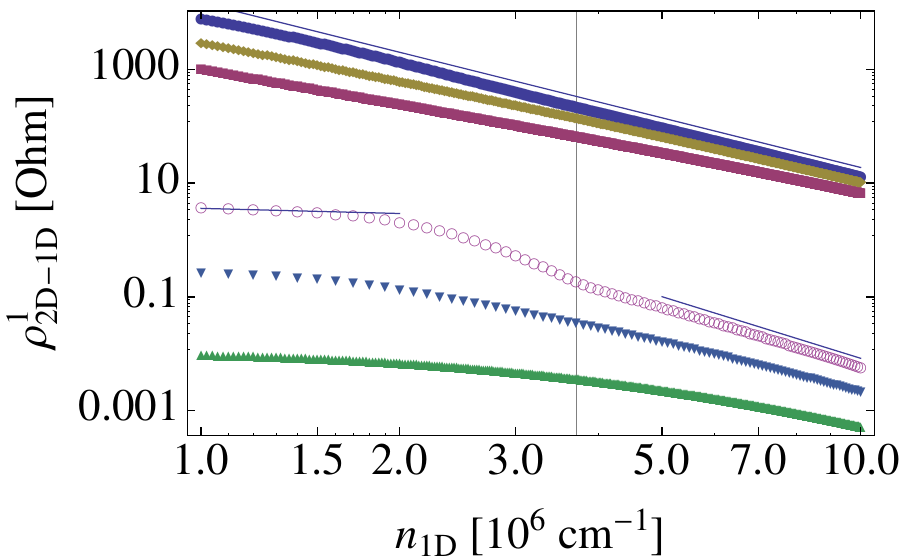}
\caption{The dependence of the drag resistivity between a semiconducting CNT and graphene on the carrier density in graphene for a fixed density $n^*_\text{1D}=3.71\times10^{6}$ cm$^{-1}$ in the CNT (left) and on the density in the CNT for a fixed density $n^*_\text{2D}=2.7\times10^{12}$ cm$^{-2}$ in graphene (right). The symbols are the log-log plots of the transresistivity calculated with (down wisp) and without (up wisp) including the screening effect for $d=30$, $10$, and $3$ nm (down up) at $T=100$ K. The radius of CNT is $R=1$ nm. The solid thin lines represent the $n_\text{2D}^{-\beta_2}$ and $n_\text{1D}^{-\beta_1}$ behaviors as a guide to the eye, respectively, on the left panel with $\beta_2=0.5$ (up wisp) and $\beta_2=1.5$, $1.45$, $0.9$ and $0.75$ (down up, down wisp) and on the right panel with $\beta_1=2.9$ (up wisp) and $\beta_1=0.3$ and $3.6$ (down wisp).}
\label{fig6}
\end{figure}

Note that the drag resistivity as function of $T$ and $d$ shows a qualitatively similar behavior for metallic and semiconducting CNTs. As seen, however, in Figs.~\ref{fig5} and \ref{fig6}, this is not the case for the drag resistivity as a function of the carrier density. In Fig. \ref{fig5} we show the  drag between a metallic CNT and graphene as a function of the carrier density in graphene. These plots show that the drag resistivity is approximately inversely proportional to $n_\text{2D}^{1.5}$ and $n_\text{2D}^{0.5}$, respectively, with and without including the screening effect. 
We find also that in this low $T$ regime $\rho^1_\text{2D--1D}\propto 1/n_\text{1D}$. This rather stable density behavior of the transresistivity in a wide range of density variations both in a metallic CNT and graphene is stipulated by the forward scattering events of Dirac electrons.

In the case of a semiconducting CNT and graphene, both forward and backward scattering processes mediate drag and their relative contribution to drag can be controlled by varying the ratio of the carrier densities. In Fig.~\ref{fig6} the vertical thin lines indicate the carrier densities $n^*_\text{2D}=2.7 \times 10^{12}$ cm$^{-2}$  (left panel) and $n^*_\text{1D}=3.71 \times 10^{6}$ cm$^{-1}$ (right panel) for matched Fermi wave vectors in graphene and CNT, $k^\text{1D}_\text{F}=k^\text{2D}_\text{F}$. These lines separate different drag scattering regimes.  At low densities in graphene, backscattering is suppressed by the presence of a graphene monolayer. Mediated by small angle scattering events, the drag resistivity decreases with an increase of $n_\text{2D}$. We observe, however, that against this overall monotonic background, the transresistivity shows a slight dip at the matching density $n_\text{2D}=n^*_\text{2D}$ ({\it cf.} the left panel in Fig.~\ref{fig6}). This new feature is due to the backward scattering channel, which opens for $n_\text{2D}>n^*_\text{2D}$.

On the right panel in Fig.~\ref{fig6} it is seen that the drag resistivity shows an upward trend as a function of $n_\text{1D}$. In this case, backward  scattering events become open for densities in CNT smaller than $n_\text{1D}<n^*_\text{1D}$, and result in a strengthened enhancement of the drag resistivity with a decrease of $n_\text{1D}$. With a further decrease of $n_\text{1D}$ the scattering phase space decreases and the dependence of the transresistivity on $n_\text{1D}$ becomes rather weak. Note that the manifestation of a sequence of different scattering regimes with variation of the carrier densities is more pronounced in samples with small values of the spacing $d$ where backscattering is significant and leads to the interplay of the small and large-angle scattering contributions to the drag resistivity.

\paragraph{Conclusions} 
We have worked out the Fermi liquid predictions for a system consisting of a CNT and monolayer graphene.  The overall physics is dominated by the dimensionality mismatch. This leads to a qualitatively novel picture of drag than that of dimensionally symmetric graphene structures.  Metallic and semiconducting CNTs show qualitatively different behavior. In particular, in structures consisting of semiconducting CNTs, the drag resistivity exhibits new features due to the competing effects of forward and backward scattering and by adjusting the charge densities one can tune the accessible scattering processes contributing to the drag resistivity.

Center for Nanostructured Graphene (CNG) is supported by the Danish National Research Foundation, Project DNRF103.

\end{document}